\documentclass[fleqn,usenatbib]{mnras}
\usepackage{newtxtext,newtxmath}
\usepackage[T1]{fontenc}
\usepackage{xcolor}
\DeclareRobustCommand{\VAN}[3]{#2}
\let\VANthebibliography\thebibliography
\def\thebibliography{\DeclareRobustCommand{\VAN}[3]{##3}\VANthebibliography}

\usepackage{graphicx}	
\usepackage{amsmath}	

\title[ Reconstructing Core-Collapse Supernovae Gravitational-Waves Waveforms]{Waveform Reconstruction of Core-Collapse Supernovae Gravitational-Waves with Ensemble Empirical Mode Decomposition}

\author[Yong Yuan et al.]{
Yong Yuan,$^{1}$
Xi-Long Fan,$^{1}$\thanks{E-mail: xilong.fan@whu.edu.cn}
 Hou-Jun L\"{u}$^{2}$, Yang-Yi ~Sun$^{3}$, Kai~Lin$^{3}$
\\
$^{1}$ School of Physics Science And Technology, Wuhan University, No.299 Bayi Road, Wuhan, Hubei, China\\
$^{2}$ Guangxi Key Laboratory for Relativistic Astrophysics, School of Physical Science and Technology, Guangxi University, Nanning, Guangxi, China\\
$^3$ School of Geophysics and Geomatics, China University of Geosciences, Wuhan 430074, Hubei, China
}

\date{Accepted 2024 February 21. Received 2024 January 25; in original form 2023 August 27}

\pubyear{2024}

\begin{document}
\label{firstpage}
\pagerange{\pageref{firstpage}--\pageref{lastpage}}
\maketitle

\begin{abstract}
The gravitational waves (GW) from core-collapse supernovae (CCSN) have been proposed as a probe to investigate physical properties inside of the supernova. However, how to search and extract the GW signals from core-collapse supernovae remains an open question due to its complicated time-frequency structure. In this paper, we apply the Ensemble Empirical Mode Decomposition (EEMD) method to decompose and reconstruct simulated GW data generated by magnetorotational mechanism and neutrino-driven mechanism within the advanced LIGO, using the match score as the criterion for assessing the quality of the reconstruction. The results indicate that by decomposing the data, the sum of the first six intrinsic mode functions (IMFs) can be used as the reconstructed waveform. To determine the probability that our reconstructed waveform corresponds to a real GW waveform, we calculate the false alarm probability of reconstruction (FAPR). By setting the threshold of the match score to be 0.75, we obtain FAPR of GW sources at a distance of 5 kpc and 10 kpc to be $6\times10^{-3}$ and $1\times10^{-2}$ respectively. If we normalize the maximum amplitude of the GW signal to $5\times10^{-21}$, the FAPR at this threshold is $4\times10^{-3}$. Furthermore, in our study, the reconstruction distance is not equivalent to the detection distance. When the strain of GW reaches $7 \times 10^{-21}$, and the match score threshold is set at 0.75, we can reconstruct GW waveform up to approximately 36 kpc.
\end{abstract}




\section{Introduction}\label{sec:intro}

The first GW signal resulting from the merger of binary black holes (BBH), known as the GW150914 event \citep{Abbott2016PhRvL}, was detected by advanced LIGO (aLIGO) in 2015 \citep{LIGO_2015CQGra}, signifying the onset of the era of GW astronomy. In 2017, the first detection of GW resulting from the merger of binary neutron stars (BNS), the GW170817 event \citep{Abbott2017PhRvL}, was detected by aLIGO and advanced Virgo \citep{Acernese_2015}, accompanied by the simultaneous observation of its electromagnetic counterparts \citep{Abbott_prl2017, Covino2017NatAs, Goldstein_apjl2017, Kasen2017Natur, Savchenko_apjl2017, Zhang2018NatCo}, heralding the era of multi-messenger GW astronomy. With the continuous upgrading and operation of GW detectors, nearly 100 GW events have already been detected by aLIGO and advanced Virgo \citep{Abbott2021GWTC3}. Furthermore, in the O4 phase, KAGRA has joined the global network of detectors \citep{Abbott2022PTEP}, leading to the anticipation of even more GW events being detected.

With the rapid progress in multi-dimensional simulations of core-collapse supernovae over the past few years, there has been a significant focus on 3D, non-rotating, neutrino-driven explosions (see Janka, Melson \& Summa 2016 \citep{Janka2016ARNPS}; M{\"u}ller 2016 \citep{Muller2016PASA} for recent reviews). Through these simulations, it has been revealed that core-collapse supernovae can potentially serve as a source of GW for aLIGO and advanced Virgo, and planned future detectors like the Einstein Telescope (ET; \citep{Punturo_2010CQGra}) and Cosmic Explorer (CE; \citep{Abbott2017CQGra}). Unfortunately, previous analysis to search GW signal from CCSN did not produce significant detection candidates.


A massive star (more than 8 $M_\odot$ at zero-age main sequence) will reach the final stage of its stellar life when the nuclear fuel has been depleted via nuclear reaction. During this phase, the core collapse is expected to occur when the mass of the core is greater than the effective Chandrasekhar mass limit \citep{Baron_apj1990, Bethe_1990}. If this is the case, two main models of CCSN are invoked to generate the GW emission, e.g., neutrino-driven mechanism \citep{Janka_2012, Bethe_1990, Bethe_apj1985}, and the magneto-rotational mechanism \citep{Scheidegger_2008, Janka_2012, Kotake_2012PTEP, Mezzacappa_2014}. The majority of GW emission is expected to emit within the first $\sim$1 s after the explosion due to convection and the standing accretion shock instability (SASI) \citep{Blondin_apj2003, Andresen_mn2017, Muller_2012, Kuroda_apjl2016, Morozova_apj2018, Yakunin_2017}.

In an ideal scenario, the recognition of the GW signal from CCSN and its extraction from the stationary and gaussian background noise of the detector, without considering transient noise or glitches \citep{Nuttall2015CQGra}, depends on the sensitivity determined by our understanding of the signal's time-frequency structure \citep{Hayama_prd2015, Gossan_prd2016}. However, the main difficulty to extract GW signal from CCSN is related to the stochasticity of its generations\citep{Morozova_apj2018}, and we can not invoke the matched filter method of searching for the generated signals from CBC to do the search for GW signal from CCSN \citep{McIver2015, Owen_prd1999}. There are many technologies to extract and reconstruct GW waveforms, such as the typical method of wavelets which is used to search and reconstructing GW waveforms generated from CBC \citep{Cornish_2015,Klimenko_2008, Lynch_prd2017}. Additionally, wavelets analysis can also be employed for reconstructing GW waveforms from CCSN \citep{McIver2015, Mezzacappa2024arXiv}. Recently years, many alternative methods have been proposed for searching and reconstructing waveforms, such as principal component analysis (PCA) \citep{Heng_2009, Rover_prd2009}, Dynamic Time Warping (DTW) \citep{suvorova_prd2019}, and Deep Learning Approach \citep{Chan_prd2020}. However, these methods are used for reconstructing or classifying GW waveforms at distances less than 100 kpc.

In this paper, we attempt to apply the ensemble empirical mode decomposition (EEMD) to extract and reconstruct the CCSN waveforms. EEMD is optimized on the basis of empirical mode decomposition (EMD) \citep{Wu2009}, which has been proposed as an adaptive time-frequency analysis method \citep{Huang_1998, Huang_1999}. The EMD can be widely applied for extracting signals from data that generated in noise with nonlinear and non-stationary processes. However, one of the major drawbacks of the EMD is the frequent appearance of mode mixing. In order to touch the problem, a new noise-assisted data analysis (NADA) method is proposed \citep{Wu2009}. Combining with NADA  and EMD, the Ensemble EMD is presented \citep{Wu2009}. It consists of decomposing an ensemble of white Gaussian noise-added signal (data) and take the mean as the final true result \citep{Wu2009}. Adding noise to the data may reduce the value of signal-to-noise ratio (SNR), but it can solve the problem of mode mixing, and will not affect the decomposition method. Thus, the motivation of this work is to extract data and reconstruct the waveform by invoking EEMD method, and analyze the signal to see how far the signal can be detected by the advanced LIGO. Finally, we calculate the false alarm probability of reconstruction.

This paper is organized as follows, in the Section \ref{sec:EEMD}, we present a brief explanation of the concept of EEMD as well as the application of EEMD for this work. In Section \ref{sec:method}, we present the data simulation and experimental results. In Section \ref{sec:simulations}, we describe the method used to obtain the simulated data. Section \ref{sec:Rec} showcased the results of decomposing and reconstructing the simulated data using the EEMD method. Finally, in Section \ref{sec:FAP}, we calculate the False Alarm Probability of reconstruction (FAPR) to analyze the probability of the reconstructed waveforms being real signals. A discussion of the results and future work is given in Section \ref{sec:dis}.

\section{EEMD Method}\label{sec:EEMD}

We  briefly  summarize  the EMD and EEMD methods. We direct the reader to \cite{Huang_1998, Huang_1999},  \cite{Flandrin2005} and \cite{Gledhill2003} for equations and related details. The EMD is proposed  as a useful adaptive time-frequency data analysis method \citep{Huang_1998, Huang_1999}, which is based on local characteristics of the original data, thus could catch nonlinear, non-stationary oscillations more effectively. By the EMD approach, the data $d(t)$ which contains signal $h(t)$ and noise $n(t)$ is decomposed in terms of Intrinsic mode functions(IMFs), $c_j$ , i.e.,

\begin{equation}
\centering
    d(t) = h(t) + n(t) = \sum_{j=1}^{k}(h_j + n_j)(t) + r_k(t)=\sum_{j=1}^{k}c_j(t) + r_k(t),
    \label{eq:emd}
\end{equation}
where $r_k(t)$ is the residue of data $d(t)$ after $k$ number of IMFs are extracted. $c_j(t)$ is the jth IMF of the decomposition and contains part of the GW data $h_j$ and the noise data $n_j$. EMD is a time-frequency method, and the IMFs' resolution is quite high in the time domain. According to Nyquist's theorem, achieving excellent resolution simultaneously in both the time domain and the frequency domain is not possible.  Consequently, the resolution in the frequency domain is not as well-defined, leading to a certain degree of frequency range overlap between adjacent IMFs. IMFs are simple oscillatory functions with varying amplitude and frequency so that they have the following properties: (1) The number of extrema and the number of zero-crossings must either be equal or differ at most by one throughout the whole length of a single IMF (even though these numbers could have a huge difference for the original data set) and (2) Regardless of the location of the data, the mean value of the envelope defined by the local maxima and the envelope defined by the local minima is zero.

One of the major shortcomings of the original EMD is the frequent occurence of mode mixing, which is defined as a single IMF either consisting of signals of widely disparate scales,  or a similar-sized signal residing in different IMF components. Mode mixing is usually a result of signal intermittency. As discussed by \cite{Huang_1998, Huang_1999}, the intermittence could not only cause serious aliasing in the time-frequency distribution, but also lead to unclear physical meaning of individual IMF.  In order to solve the problem of scale separation, a new noise-assisted data analysis (NADA) method is proposed, the Ensemble EMD, which defines the true IMF components as the mean of an ensemble of experiments, each of which consists of the signal plus a white Gaussian noise of finite amplitude. Furthermore,  noise could contribute to data analysis in the EMD \citep{Flandrin2005,Gledhill2003}.

To improve the precision of measurements, the ensemble mean is a powerful method, where data are collected by individual observations, each of which contains different gaussian white noise. To generalize this ensemble method, gaussian white noise is introduced to a single data set, $d(t)$, it's like a physics experiment that can be repeated many times, but is making different observations, $s_i(t)$. The added gaussian white noise is treated as the possible random noise that would be encountered in the detection process. In this case, the $i$th observation will be:

\begin{equation}
\begin{aligned}
\centering
s_i(t) & = d(t) + N_i(t)= \sum_{j=1}^{k}(h_{i,j} + n_{i,j} +  N_{i,j})(t) + r_{i,k}(t)\\
 &=\sum_{j=1}^{k}b_{i,j}(t) + r_{i,k}(t),
\label{eq:eemd}
\end{aligned}
\end{equation}

\begin{equation}
\centering
b_j(t) = \frac{\sum_{i=1}^{I}b_{i,j}(t)}{I}
\label{eq:cj}
\end{equation}
In the case of only one detection, each multiple-observation ensemble is simulated by adding different implementations(not arbitrary but different ) of gaussian white noise, $N_i(t)$ ($N_i(t)\ll n(t)$), to that single detection, as shown in Eq.(\ref{eq:eemd}). $b_{i,j}(t)$ contains part of the gravitational wave signal $h_{i,j}(t)$, instrument noise $n_{i,j}(t)$ and gaussian white noise $N_{i,j}(t)$. $b_{j}(t)$ is the sum and average of the components under the same index as the final result. $I$ is the total number of repetitions. Although the addition of noise may lead to a relatively small signal-to-noise ratio (SNR), the added white Gaussian noise will provide a relatively uniform reference scale distribution to facilitate EMD; therefore, the low SNR does not affect the decomposition method but enhances it to avoid the mode mixing.

The proposed EEMD is developed as follows: (1) add a gaussian white noise series to the targeted data; (2) decompose the data with added noise into IMFs; (3) repeat step 1 and step 2 again and again, but use a different white gaussian noise series each time and (4) obtain the (ensemble) means of corresponding IMFs of the decomposition to represent the final result.

The result of the decomposition using the EEMD is that the added white Gaussian noise time series cancel each other in the final mean of the corresponding IMFs; the mean IMFs effectively function as a collection of bandpass filters, thus significantly reducing the chance of mode mixing and preserve the dyadic characteristics.

\section{Simulations and Results}
\label{sec:method}

\subsection{Data simulations}
\label{sec:simulations}

In this section, we describe the stages of the testing process to aimed at evaluating the effectiveness of the EEMD method in improving waveform reconstruction. As a preliminary attempt in this paper, we focus solely on a single  aLIGO detector to verify the feasibility of the EEMD method. 

As a first step, we utilize simulated waveforms obtained from various sources in the literature. The magnetorotational CCSN signals are taken from \citep{Abdikamalov2014PhRvD, Dimmelmeier2008PhRvD, Richers2017PhRvD}\footnote{\url{https://stellarcollapse.org/gwcatalog.html}, \url{https://zenodo.org/record/201145.}}, with the GW waveforms generated by \cite{ Dimmelmeier2008PhRvD} based on 2-D simulations. For the neutrino-driven mechanism, we employ the waveforms from \citep{Yakunin_2017,  Ott2013ApJ, Andresen2019MNRAS, Kuroda2017ApJ, Muller2012AA, powell_mn2019, Radice2019ApJL, powell_mn2020}\footnote{\url{https://wwwmpa.mpa-garching.mpg.de/ccsnarchive/data/Andresen2019/}, \url{https://www.astro.princeton.edu/~burrows/gw.3d/}}. These simulations encompass a wide range of progenitor masses from 9 $M_\odot$ to 60 $M_\odot$ are shown in Table \ref{tab:waveforms}. Given that these waveforms are generated under different conditions such as varying distances, sampling rates, and durations, it becomes necessary to normalize them before utilizing them to generate the time series. To achieve this, we scale the amplitudes of the waveforms by relocating the sources to a distance of 10 kpc from Earth. Additionally, we ensure uniformity in the sampling rates by down-sampling all waveforms to a preselected rate of 4096 Hz. It should be noted that some of the waveforms utilized in this study are generated under axisymmetric simulations. In such cases, the waveforms are solely described by the polarization $h_+(t)$, while the corresponding $h_\times(t)$ components are defined as vectors of zeros.

The next step is to simulated the data observed by aLIGO according to Eq (\ref{eq:data}). 
\begin{equation}
d(t)=F_{+}(\alpha, \delta)h_{+}(t)+F_{\times}(\alpha, \delta)h_{\times}(t)+n(t) = h(t) + n(t),
\label{eq:data}
\end{equation}
where $n(t)$ is created using the zero detuning, high power, noise power spectral density (PSD) for aLIGO at design sensitivity \citep{Abbott2020LRR}, $h_{+}(t)$ and $h_{\times}(t)$ are strains in two independent polarization modes and $F_{+}(\alpha, \delta)$ and $F_{\times}(\alpha, \delta)$ are the antenna patterns function ($\alpha$ is right ascension and $\delta$ is declination). A random location of ($\alpha$, $\delta$) in the sky is selected from a uniform distribution on ($\alpha$, $\sin \delta$). Owing to the short duration of the simulated signals (about 1 second), it is appropriate to assume that both antenna pattern functions are time-independent and thus real constants.

\begin{table*}
\centering
\caption{The mass ranges and mechanisms of the progenitors associated with the simulated waveforms used in this work. The first column corresponds to the studies. Mechanism indicates the explosion mechanism for the waveforms, denoted by M for the magnetorotational mechanism and N for the neutrino-driven mechanism. M1, M2, N1, and N2 represent the waveforms we have chosen for case study. Mass refers to the mass of the progenitors in solar mass units within the simulations. No. means the number of waveforms available from the study. All mentioned masses correspond to the stellar masses at zero age unless otherwise stated.}\label{tab:data}
\begin{tabular}{llll}
\hline
\hline
             & Mechanism & Mass ($M_{\odot}$)                       & No.  \\ \hline
Abdikamalov \citep{Abdikamalov2014PhRvD}  & M (M1)        & 12.0                          & 92   \\ 
Dimmerlmeier \citep{Dimmelmeier2008PhRvD} & M        & 11.2, 15.0, 20.0, 40.0        & 136  \\ 
Richers \citep{Richers2017PhRvD}      & M (M2)       & 12.0                          & 1824 \\ 
Andresen \citep{Andresen2019MNRAS}     & N        & 11.2, 15.0                    & 6    \\ 
Kuroda \citep{Kuroda2017ApJ}       & N        & 11.2, 15.0                    & 2    \\ 
Muller \citep{Muller2012AA}       & N        & 15.0, 20.0                    & 6    \\ 
Ott \citep{Ott2013ApJ}          & N        & 27                            & 8    \\ 
Powell \citep{powell_mn2019}       & N        & 18.0                     & 1    \\ 
Powell \citep{powell_mn2020}       & N (N1)       &  39.0                     & 1    \\ 
Radice \citep{Radice2019ApJL}       & N        & 9, 10, 11, 12, 13, 19, 25, 60 & 8    \\ 
Yakunin \citep{Yakunin_2017}      & N (N2)       & 15                            & 1    \\ 
\hline
\hline\\

\end{tabular}

\label{tab:waveforms}
\end{table*}

Before decomposing the data using the EEMD method, preprocessing is necessary. This involves conducting a spectral analysis, whitening the colored noise of the detector to approach gaussian white noise, and applying bandpass filtering with a frequency range of [20, 2000] Hz. We present the simulated data after whitening and bandpass filter as $\tilde{d}(t)$. This process helps to make the noise approach gaussian white noise.In addition, we selected simulated data with a time window of 1 second for further processing. During the process of decomposing the data using the EEMD method, it is necessary to introduce additional gaussian white noise into the data. To ensured that the added Gaussian white noise does not affect the detection and reconstruction of GW, we set the mean of the additional gaussian white noise to 0 and the standard deviation to 0.001$\times|\text{max}(\tilde{d}(t))-\text{min}(\tilde{d}(t))|$ . Additionally, we conduct 1000 trials in the EEMD method, striking a balance between stability in the decomposition results and efficient utilization of computational resources. 

\subsection{Decomposition and Reconstruction}
\label{sec:Rec}

\begin{figure}
\begin{center}

    \includegraphics[width=8.5cm]{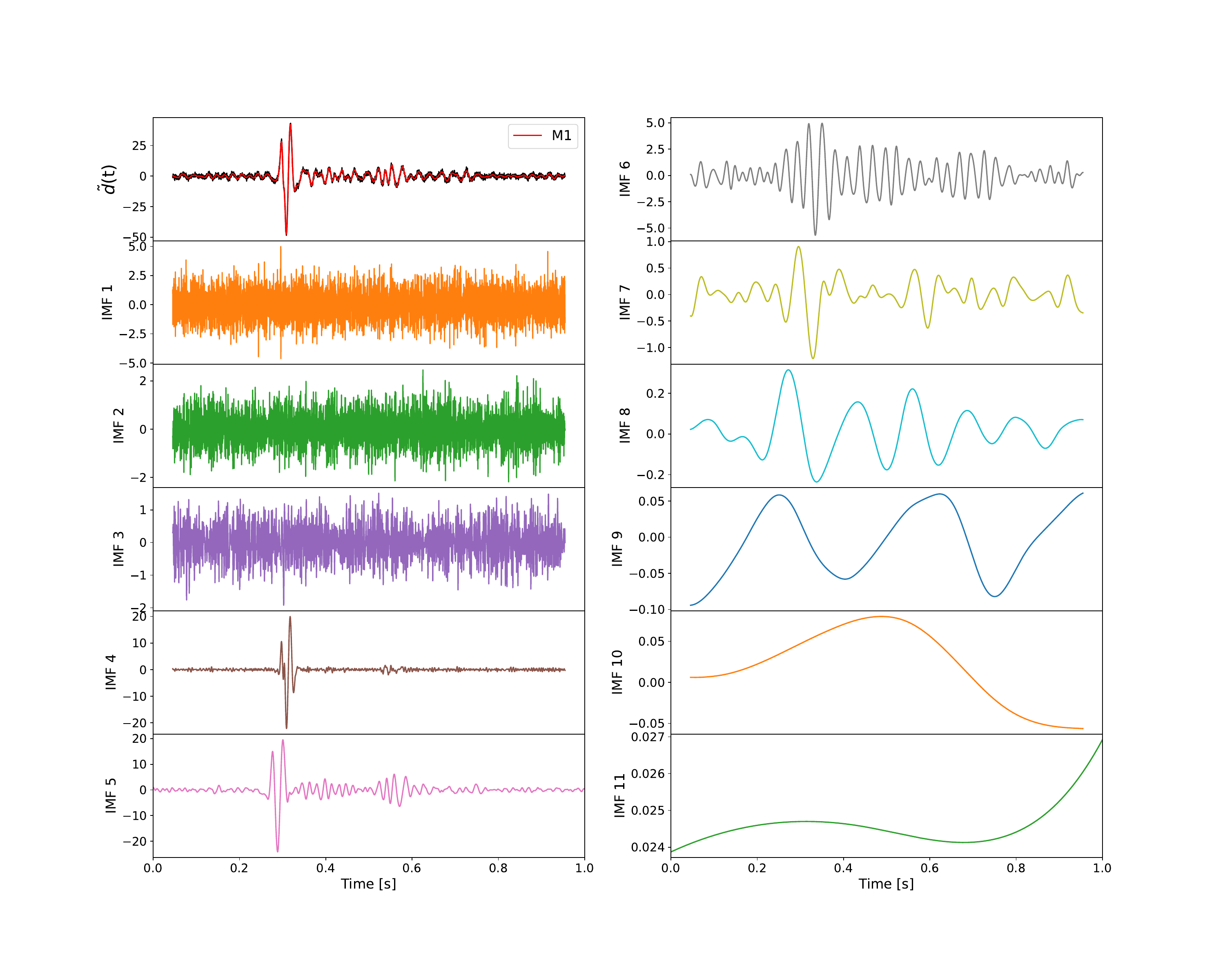}

    \includegraphics[width=8.5cm]{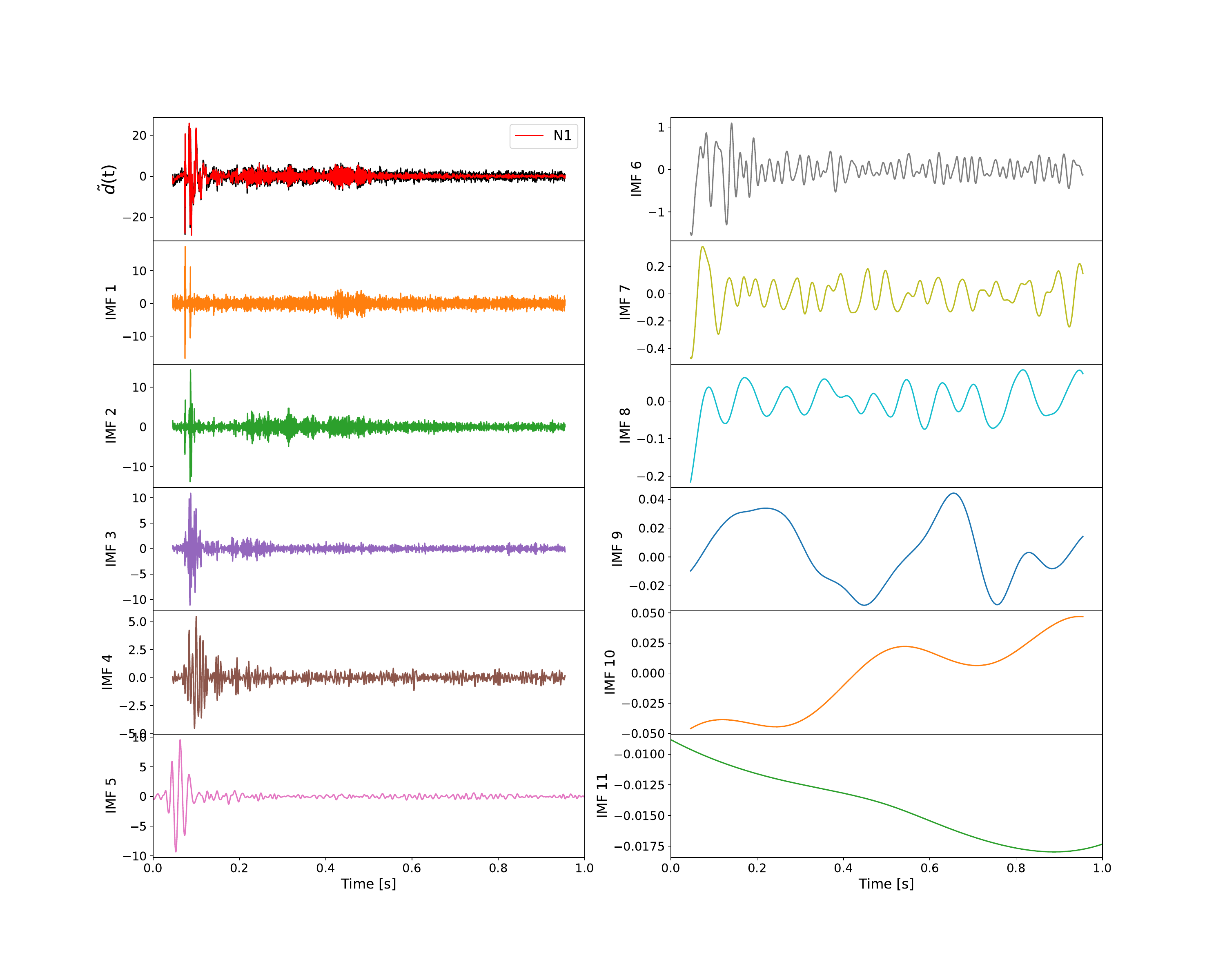}
	\caption{Using the EEMD method to decompose the simulated data after whitening and bandpass filter (black line) injected with the GW waveform (red line) at a distance of 5 kpc, resulting 11 IMF components. The frequencies of these components decrease sequentially from IMF1 to IMF11. The top graph depicts the injection of the M1 waveform, while the bottom graph illustrates the injection of the N1 waveform.}
	\label{fig:Abd_gw_eemd_dhp}
\end{center}
\end{figure}

We have selected two waveforms from Table \ref{tab:data} that exhibit significant difference in both time-domain and frequency-domain structures as case studies for illustrating EEMD method. One waveform is obtained from \cite{Abdikamalov2014PhRvD}, denoted as M1, is generated by magnetorotational mechanism. This waveform possesses a relatively simple time-domain structure, consisting of two consecutive peaks, with a frequency distribution in the range of 20-100 Hz in the frequency domain (as indicated by the red curve in the first row of the top-left side of Figure \ref{fig:Abd_gw_eemd_dhp}). The other waveform obtained by \cite{powell_mn2020}, denoted as N1, is generated by neutrino-driven mechanism, including several prominent peaks along with additional smaller but rapid oscillatory structures. In the frequency domain, this waveform's frequency distribution is relatively broad, covering a range from 20-1200 Hz (as indicated by the red curve in the first row of the bottom-left side of Figure \ref{fig:Abd_gw_eemd_dhp}). Using the simulated data from these two waveforms, we demonstrated the decomposition process of data using the EEMD method. The results are shown in Figure \ref{fig:Abd_gw_eemd_dhp}. The simulated data are decomposed into IMFs from high to low frequency, in which the left first row is the simulated data contained signal and noise. We can found that the amplitude of these components are so low that can be neglected after IMF6 from Figure \ref{fig:Abd_gw_eemd_dhp}. Furthermore, the main components of the GW waveform are not confined to any specific IMF.

By applying the EEMD method to decompose the $\tilde{d}(t)$, we obtain many IMF components. It is crucial for us to determine the optimal number of components to be employed in reconstructing the GW waveform. Furthermore, our objective is to assess the fidelity of the fidelity of the reconstruction by evaluating the degree of concordance between the reconstructed GW waveform and the injected GW waveform. Therefore, we use the match score ($\eta$) as a metric to measure the accuracy of the reconstruction \citep{suvorova_prd2019}:
\begin{equation}
\eta = \frac{(\tilde{h}|\bar{h})}{\sqrt{(\tilde{h}|\tilde{h})(\bar{h}|\bar{h})}},
\label{match_score}
\end{equation}
where $\tilde{h}$ is simulated signal after bandpass filtering, $\bar{h}$ is the reconstructed waveform depending on the adopted number of IMFs, and $(a|b)$ denoted the inner product\footnote{
In \cite{suvorova_prd2019}, the inner product includes a noise weighting term, equivalent to whitening the data. In our processing, we have already whitened the data, so there is no need to apply noise weighting to the inner product here.}.

\begin{figure}

    \includegraphics[width=8cm]{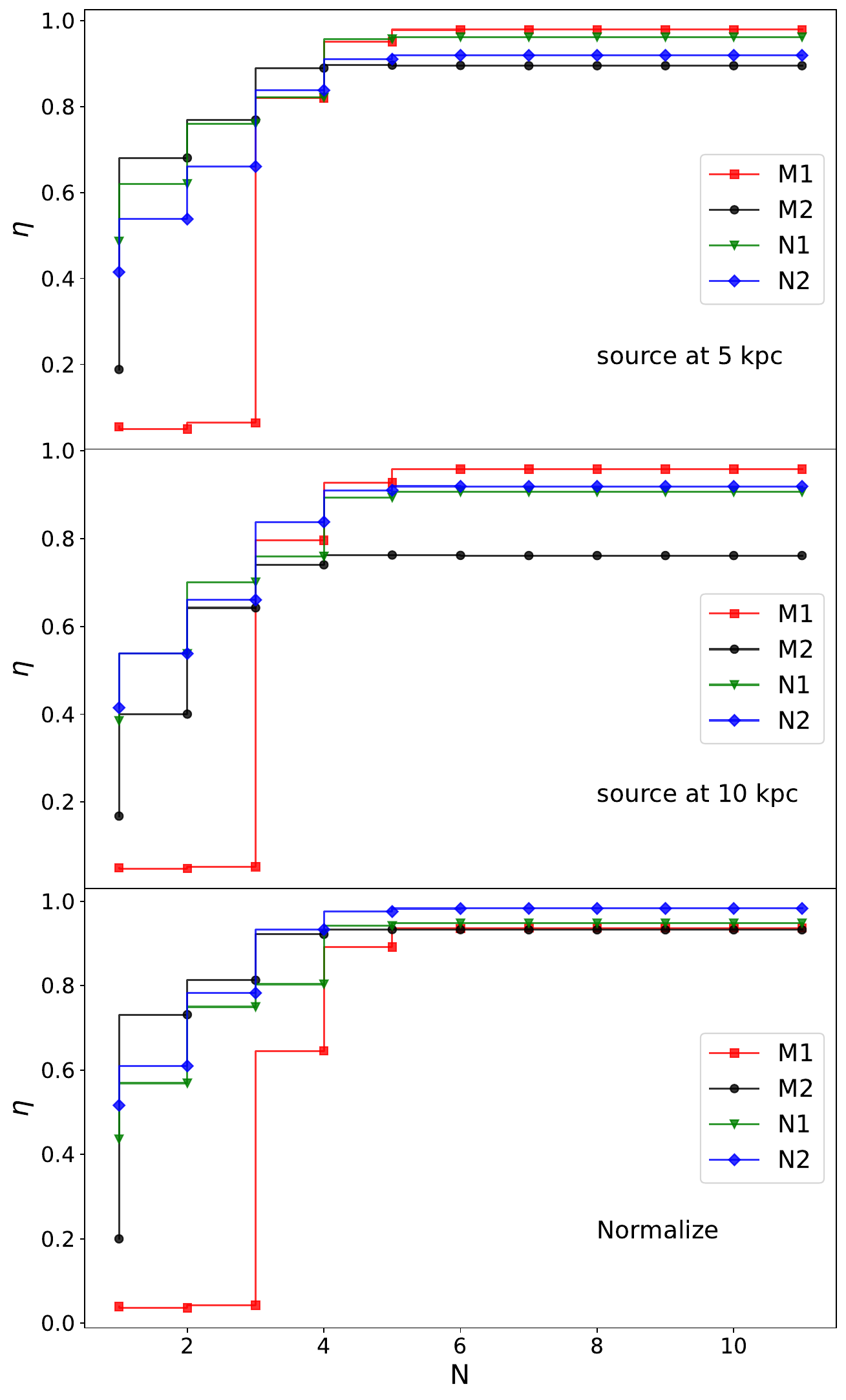}
        
\caption[]{The relationship between the match score and the number of IMFs. The top graph presents four waveforms at 5kpc, the middle graph at 10kpc and the bottom graph depicts the maximum amplitude of four waveforms are normalized to $5\times10^{-21}$. The waveforms shown by M1 and M2 are generated by magnetorotationa mechanism, and the waveforms shown by N1 and N2 are generated by neutrino-driven mechanism. \label{fig:Match}}
\end{figure}

We conduct on multiple sets of simulated data to evaluate the match scores as a function of the number of IMF components. In this section, in addition to using the previously mentioned M1 and N1 waveforms, we also utilized two additional waveforms generated by magnetorotational mechanism and neutrino-driven mechanism, denoted as M2 and N2, respectively. Specifically, M2 is obtained from \cite{Richers2017PhRvD}, and N2 is obtained from \cite{Yakunin_2017}. These simulated data were employed to demonstrate the performance and application of the EEMD method under different mechanisms. The results illustrate the evolution of match scores with the number of IMF components at 5 kpc and 10 kpc. To eliminate the influence of GW waveform amplitudes, we normalized all the GW data ($h(t)$) by setting their maximum amplitude to $5\times10^{-21}$ (corresponding to the maximum amplitude of N1). The results are shown in Figure \ref{fig:Match}. Next, we apply the EEMD method to decompose $\tilde{d}(t)$ and analyze the evolution of match scores with the number of IMF components. From the results in Figure \ref{fig:Match}, it is evident that the match score is already very close to its maximum value when comparing the sum of the first five IMF components to the injected waveform. Nevertheless, to preserve essential structural features in the waveform and prevent their loss, we ultimately choose the sum of the first six IMF components as the reconstructed waveform, $\bar{h} = \sum_{j=1}^6 b_j(t)$. We also find that the match score of GW obtained from the magnetorotational mechanism is not necessarily larger or smaller than the match score obtained from the neutrino-driven mechanism. Therefore, we cannot solely rely on the match score to distinguish the explosion mechanism of GW. Furthermore, it is found from Figure \ref{fig:Match} that the first three IMFs of M1 have relatively low match scores, around 0.1. It is only when the fourth IMF is included that the match score increases significantly, exceeding 0.6, which is consistent with the results shown in top graph of Figure \ref{fig:Abd_gw_eemd_dhp}. This is because the first three IMFs of M1 mainly consist of high-frequency noise components, which do not contribute significantly to the match score. Compared to the M1 waveform, the other three waveforms have higher-frequency components. Therefore, as the number of IMF components increases, the matching score will also increase until it reaches a stable state.

\begin{figure}
\centering
    \includegraphics[width=8cm]{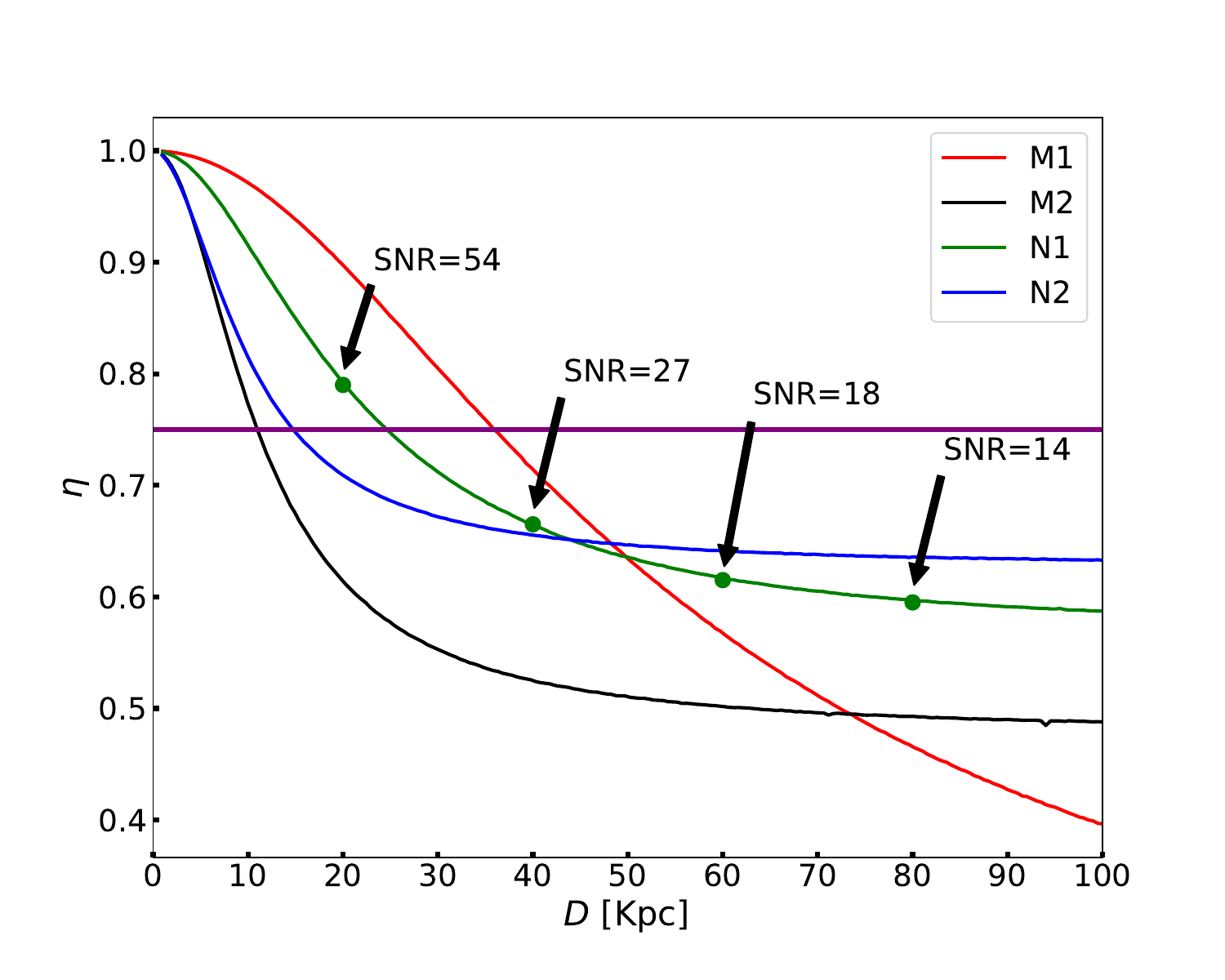}
    \caption[]{The evolution of the match score with distance. The waveforms used are consistent with those in Figure \ref{fig:Match}. The red, green, black and blue lines represent the match score between the reconstructed waveform and the injected waveform at different distances for each waveform, respectively. The purple horizontal line represents the threshold, which is 0.75. Furthermore, the green dots on the graph represent the match score we calculated for the N1 waveform at 20 kpc, 40 kpc, 60 kpc, and 80 kpc, with the corresponding SNR marked on the graph at those positions.
\label{fig:dis}}
\end{figure}

The next step, we analyze the ability of the EEMD method to reconstruct GW waveforms at different distances. We simulate GW observation data at various distance and subsequently apply the EEMD method to decompose the simulated observation data, thereby reconstructing the GW waveforms. Furthermore, We calculate the match score between the reconstructed waveforms and the injected waveforms. We continue to use the four models shown in Figure \ref{fig:Match} for demonstration, and the results are presented in Figure \ref{fig:dis}. From the results in Figure \ref{fig:dis}, it is evident that the match scores decrease as the distance increases. This is attributed to the diminishing amplitude of GW with increasing distance. Consequently, the components obtained through the EEMD method decomposition contain more noise, leading to a decrease in the match score. However, the decreasing trend gradually becomes smoother in the end, primarily because the extracted data and injected waveform partially match, and their variation does not solely depend on the distance.

In order to determine whether the GW waveform has been successfully reconstructed, it is necessary to establish a threshold for the match score. In this study, we set the threshold at 0.75. If the match score is greater than or equal to 0.75, it indicates that the GW waveform has been extracted and reconstructed from the observed data. Conversely, if the match score is less than 0.75, it implies that the reconstructed waveform predominantly consists of noise. From Figure \ref{fig:dis}, it is evident that the successful reconstruction distance for the four waveforms are all less than 50 kpc, with the farthest distance reaching up to 36 kpc. However, according to \cite{powell_mn2020} study, the waveform of N1 can be detected at distance of up to 100 kpc. We conducte a verification test at a distance of 100 kpc, and signal-to-noise ratio (SNR) exceeds 8, indicating that the GW signal of N1 can indeed be detected by aLIGO (see details in Figure \ref{fig:dis}). This suggests that the detection distance is not equivalent to the reconstruction distance. Furthermore, from Figure \ref{fig:dis}, it can be observed that in the final stable state (where match score do not vary with distance), compared to the N1 and N2 waveforms, the match score for the M1 and M2 waveforms are lower. The reason for this result is that, compared to M1 and M2, N1 and N2 exhibit more complex structures in both the time and frequency domains, resembling random noise. Specifically, the waveforms of N1 and N2 are decomposed into more IMFs. As a consequence, when the distance exceeds 50 kpc and noise dominates the reconstructed waveform, N1 and N2 show a higher degree of similarity to the noise. This result is also consistent with the distribution of match score in Figure \ref{fig:FAR_dis_case} when no GW  waveform was injected.

\begin{figure*}
    \includegraphics[width=8cm]{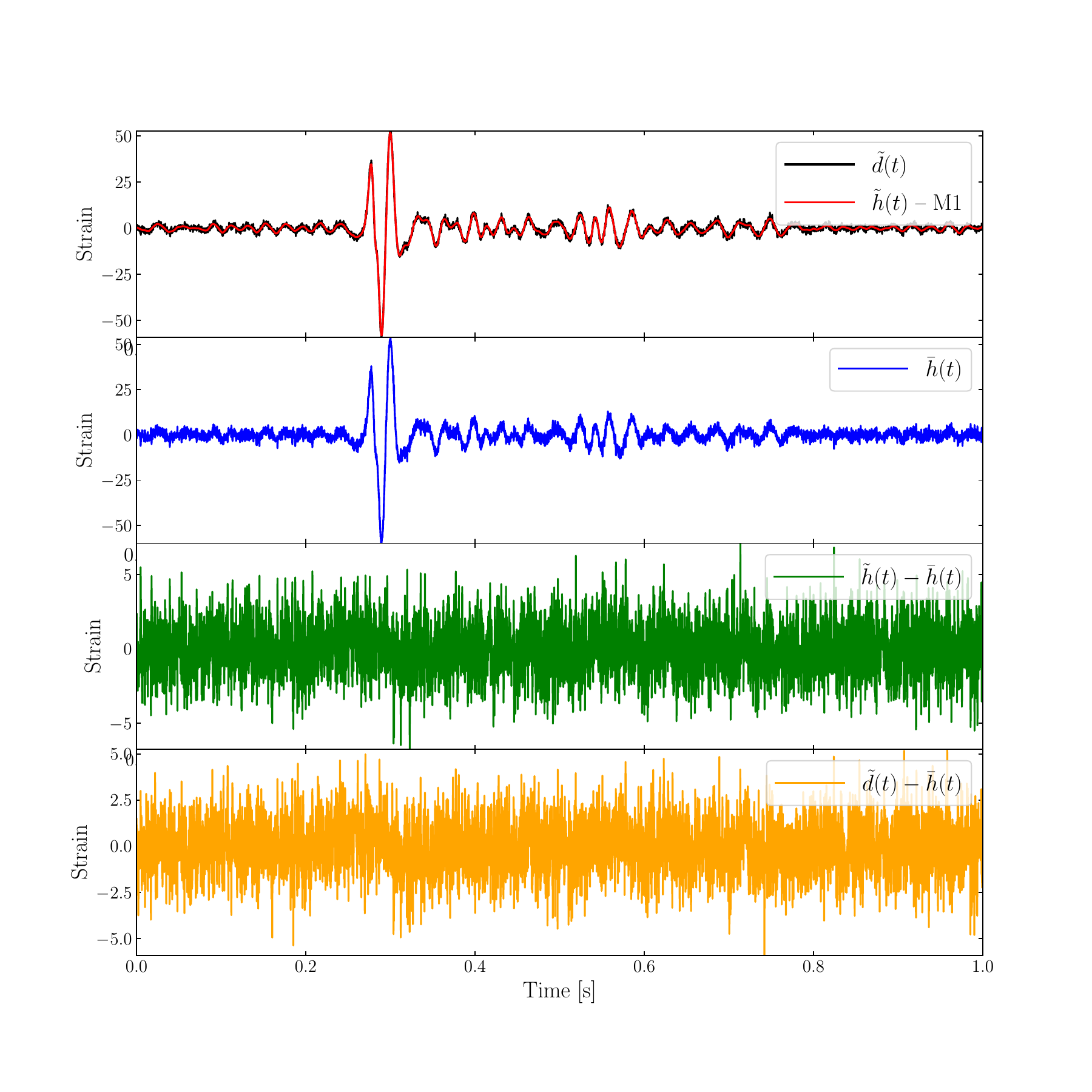}
    \includegraphics[width=8cm]{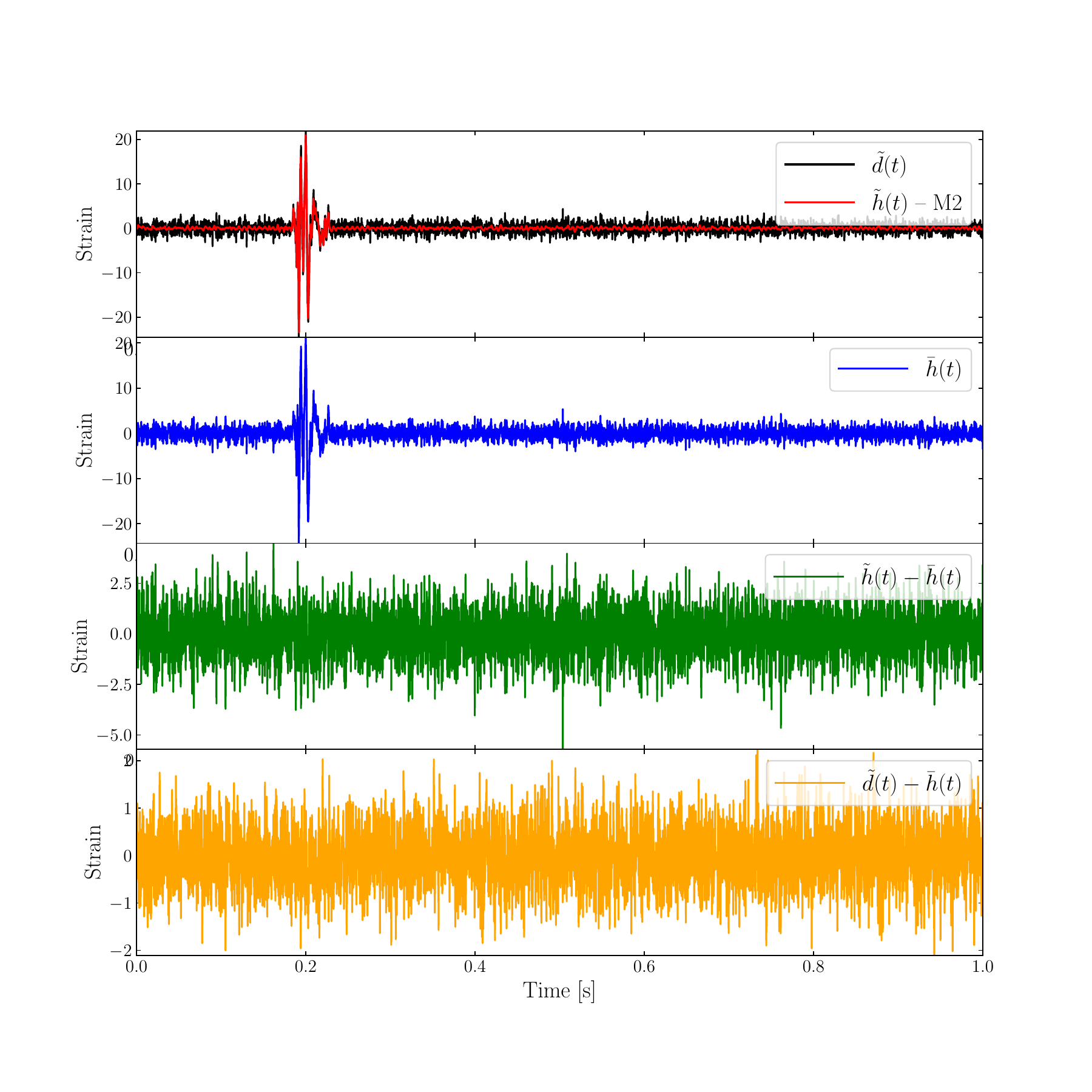}
    
    \includegraphics[width=8cm]{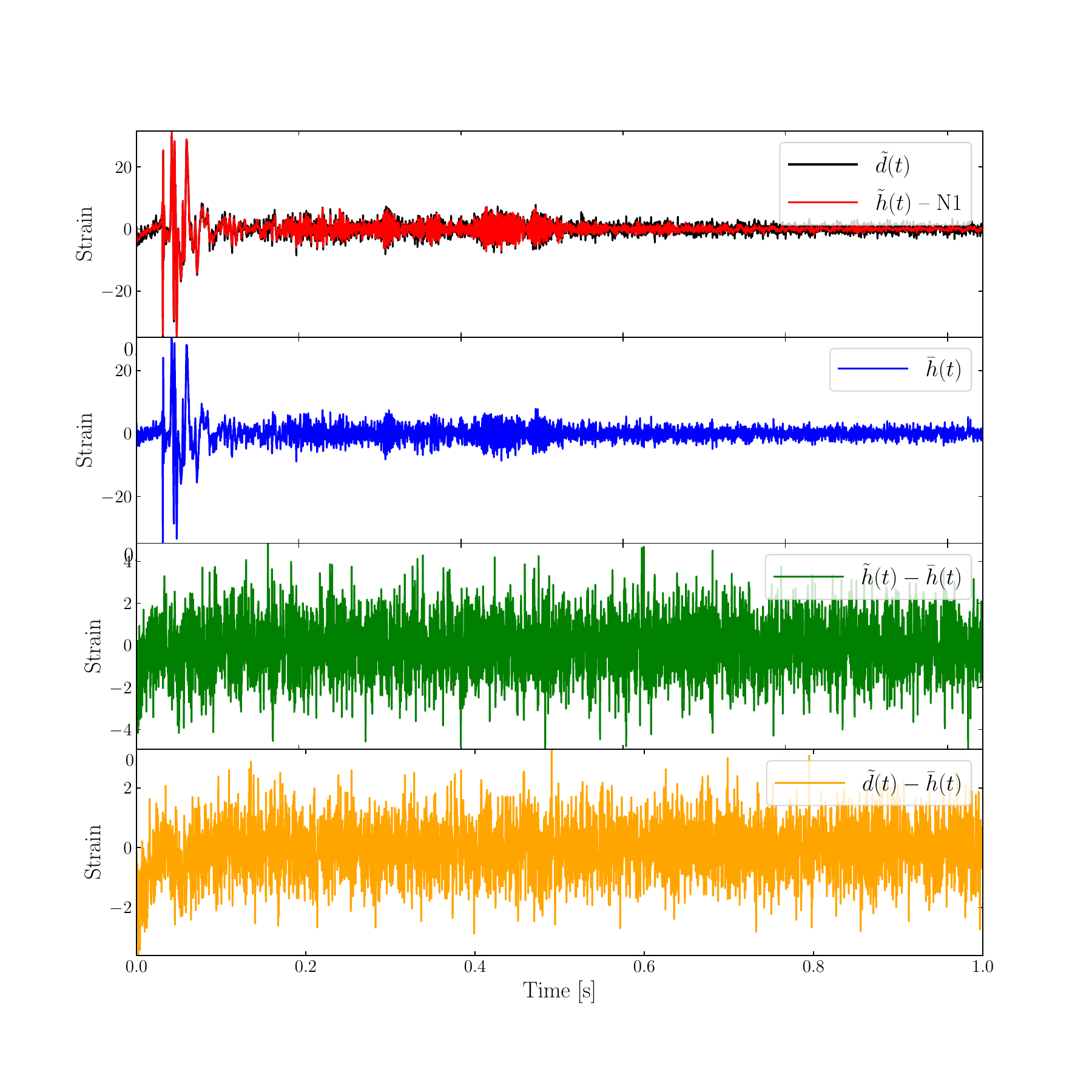}
    \includegraphics[width=8cm]{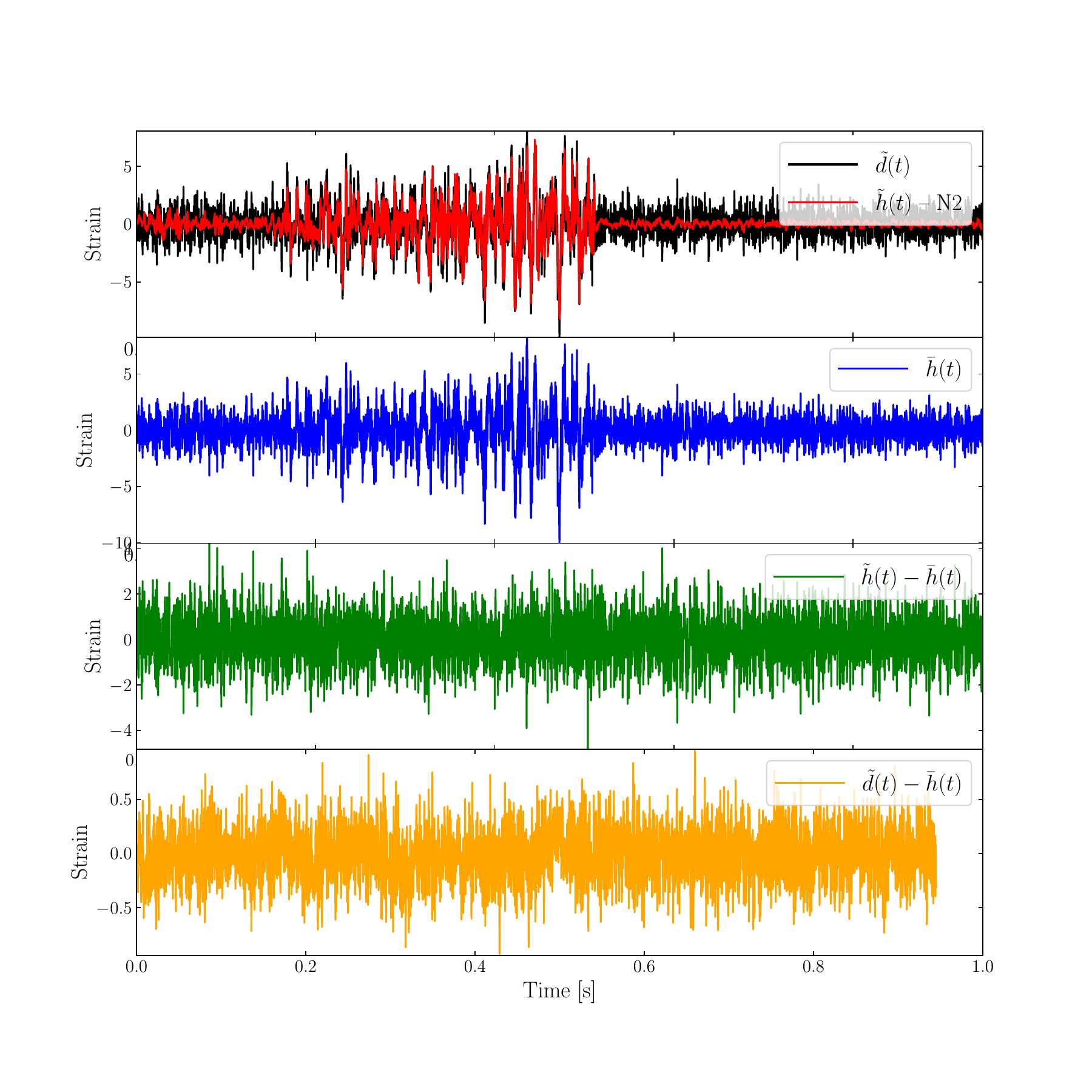}

\caption[]{The results of decomposing and reconstructing the four waveforms at a distance of 5 kpc using the EEMD method. The waveforms used are consistent with those in Figure \ref{fig:Match}. In each image, the top panel represents the whitened and filtered injected waveform (red line) and the simulated data (black line), the second panel represents the reconstructed waveform (blue line), the third represents the residual between the whitened and filtered injected waveform and the reconstructed waveform (green line), and the bottom panel represents the residual between the whitened and filtered simulated data and the reconstructed waveform (orange line). \label{fig:rec_h}}
\end{figure*}

We present the results of reconstructing the four waveforms at a distance of 5 kpc, as shown in Figure \ref{fig:rec_h}. Through the displays in the second row (reconstructed waveforms) and the third row (residuals between injected waveforms and reconstructed waveforms) of each subplot in Figure \ref{fig:rec_h}, we are able to effectively reconstruct the principal characteristics of the injected waveforms, even though there might be some noise components present in the reconstructed waveforms. This phenomenon arises due to the fact that during the data decomposition and reconstruction process using the EEMD method, each individual component encompasses not only the signal component but also a fraction of noise components. Further observation of the fourth row (residuals between simulated data and reconstructed waveforms) in each subplot of Figure \ref{fig:rec_h} indeed indicates that the EEMD method is capable of extracting GW signals from the noise.

 \subsection{False alarm probability of reconstruction }
 \label{sec:FAP}
 
 In addition, we also calculate the match score between the reconstructed waveform, obtain without injecting any GW waveforms, and the injected GW waveforms. This is done to verify the accuracy of the reconstructed waveform and prevent the possibility of mistaking noise for a signal. We calculate the threshold $\eta$ from the cumulative distribution of the match score for noise-only realization $P_{\text{noise}}$, in order to obtain the FAPR as shown in Eq.(\ref{eq:FAP}):
\begin{equation}
1-P_{\text{noise}}(\eta) \leq \text{FAPR},
\label{eq:FAP}
\end{equation}

\begin{figure}
\centering
    \includegraphics[width=8cm]{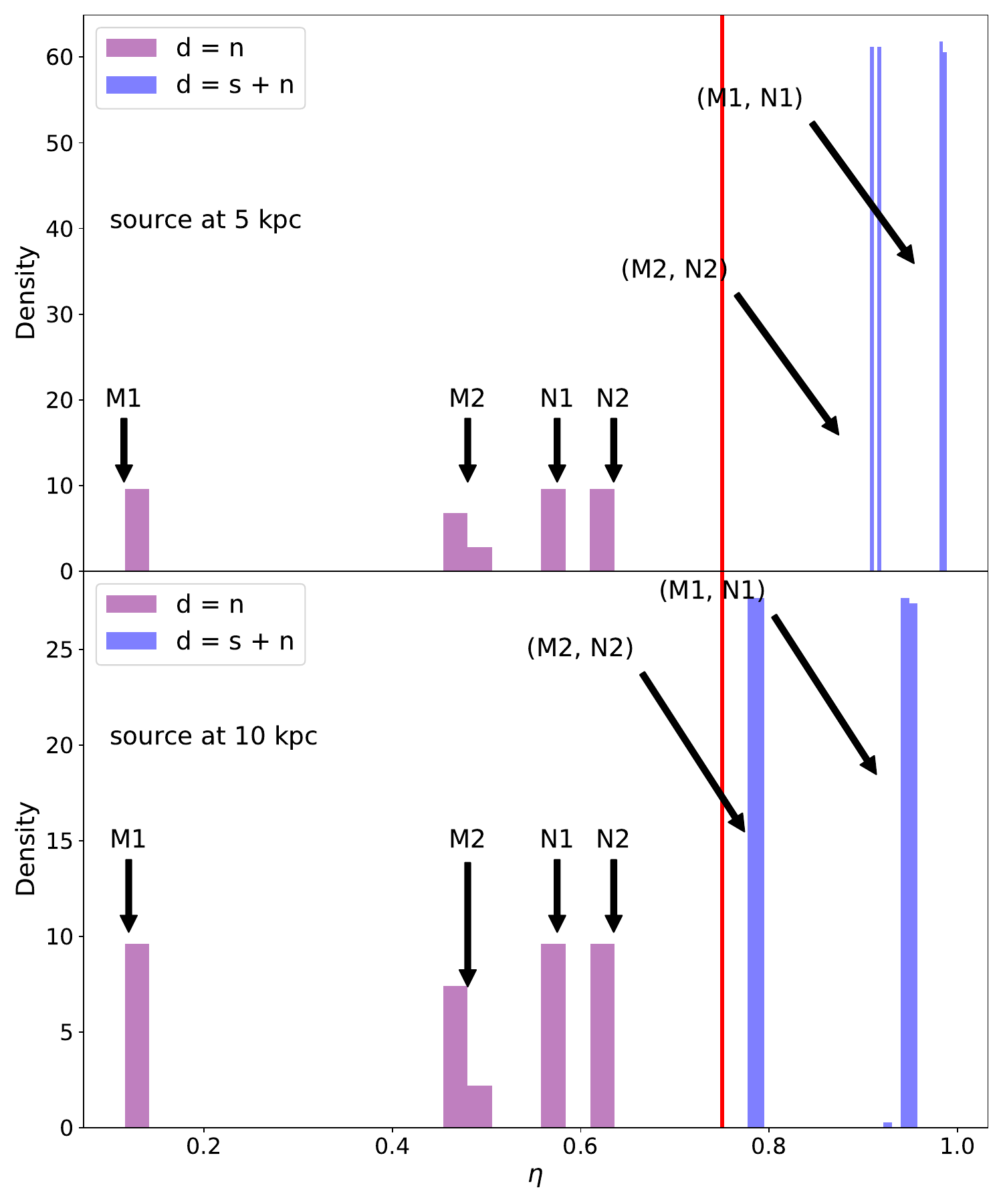}
\caption{
Histogram of the match score for the simulated data, including the distribution of match scores with waveform injection (blue) and without waveform injection (purple). The injected waveforms include M1, M2, N1 and N2, with equal occurrences for each waveform. The red vertical line represents the threshold of 0.75. The top graph represents the match score distribution for the sources at 5 kpc, and the bottom graph represents the match score distribution for sources at 10 kpc. \label{fig:FAR_dis_case}}
\end{figure}

At 5kpc, we use the aforementioned 4 waveforms to generate 100 simulated data sets for each waveform, resulting in a total of 400 simulated data sets with waveforms injected. We calculate their distributions of match score. Additionally, we generate 400 sets of pure noise data, apply the EEMD method for decomposition and reconstruction, then averaged them into 4 groups, and match each group with the 4 waveforms individually to obtain the distribution of match scores without waveform injection. Similarly, at 10kpc, we perform the same procedure to obtain the distributions of matching scores with and without waveform injection. The results are shown in Figure \ref{fig:FAR_dis_case}. When the threshold $\eta = 0.75$, the FAPR at both 5 kpc and 10 kpc is less than $2.5\times10^{-3}$. This result indicates that the match scores obtained through the EEMD method can effectively distinguish between GW signals and pure noise, confirming the reliability and effectiveness of this method in reconstructing GW signals. From the results in Figure \ref{fig:FAR_dis_case}, we find that in the distribution of match score without waveform injection (purple), approximately 1/4 of the results are distributed to the left of 0.2. This is mainly due to the lower match score between the reconstructed waveform without injected waveforms and the M1 waveform. Compared to the other three waveforms, the M1 waveform has relatively simpler components and larger amplitudes, leading to a lower degree of match with the noise and, consequently, resulting in lower match score. In addition, in the match score distribution of the injected waveforms (blue), we also find two peaks. The smaller peak is mainly composed of the match score results for M2 and N2 waveforms, as M2 waveform has a relatively simpler structure and N2 waveform has a relatively smaller amplitude. As a result, the match score for these waveforms are lower compared to the other waveforms. Furthermore, based on the results, we can still conclude that the match score is currently not suitable for directly classifying GW waveforms.

\begin{figure}
\centering
    \includegraphics[width=8cm]{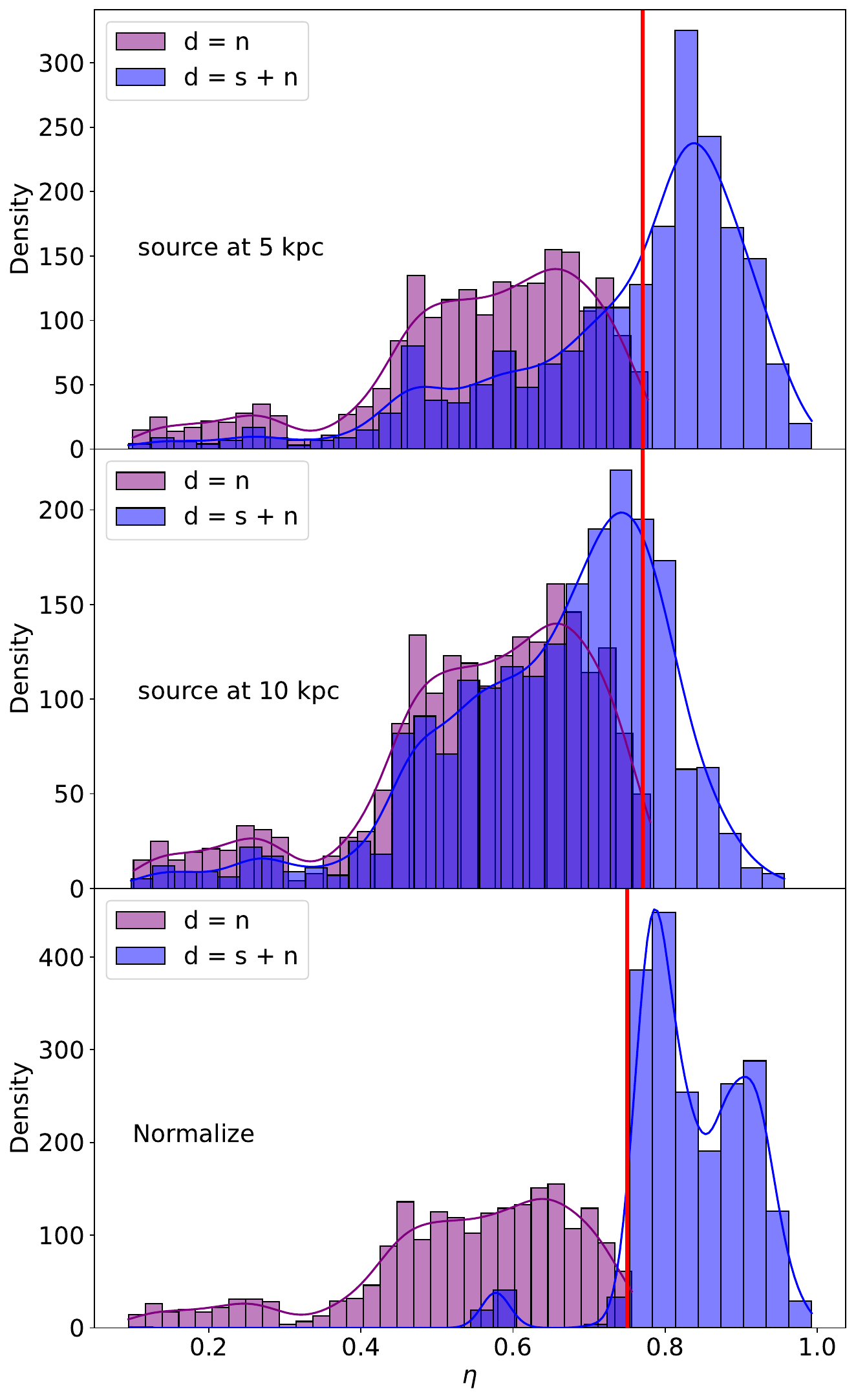}
\caption{
Histogram of the match score for the simulated data, including the distribution of match scores with waveform injection (blue) and without waveform injection (purple). The injected waveforms consist of the 2085 waveforms listed in Table \ref{tab:waveforms}, with equal number of injections for each waveform. The red vertical line represents the threshold of 0.75. The curves on the histogram were obtained using the kernel density estimation. The top graph represents the match score distribution for the sources at 5 kpc and FAPR = $6\times 10^{-3}$,  the middle graph  depicts the match score distribution for sources at 10 kpc and FAPR = $1\times 10^{-2}$. Meanwhile, the bottom graph illustrates the maximum amplitude of the injected waveforms are normalized to $5 \times 10^{-21}$ and FAPR = $4\times 10^{-3}$.
\label{fig:FAR_dis}}
\end{figure}

Finally, we analyze the 2085 waveforms from Table \ref{tab:waveforms}. At 5kpc, each waveform was injected twice, resulting in a total of 4170 simulated data. We obtain the match score with waveform injection from these data. Additionally, we generate 4170 pure noise signals without waveform injection and obtained the match score without waveform injection. Similarly, we performe the same procedure at 10 kpc and obtained the results shown in the top and middle graph of Figure \ref{fig:FAR_dis}. The curves on the histogram in Figure \ref{fig:FAR_dis} are obtained using the kernel density estimation. When the threshold $\eta=0.75$, the FAPR at 5 kpc and 10 kpc are $6\times10^{-3}$ and $1\times10^{-2}$ respectively. From the results in the top and middle graph of Figure \ref{fig:FAR_dis}, it can be observed that the FAPR at 5kpc is lower than at 10kpc. Additionally, a significant portion of the data injected with GW waveforms yields match score below the threshold we set. This is primarily due to two reasons: first, these data are not in the optimal sky region for detector response, and second, the injected waveforms have inherently small strain ($\text{max}(h(t)) < 8 \times 10^{-22}$). The combined effect of these factors results in many simulated data with injected GW waveforms not being successfully reconstructed. Additionally, near a match score of 0.2, we find a small peak. This is because some waveforms have structures similar to the N1 waveform, but their amplitudes are very small,  e.g.$\sim 1 \times 10^{-22}$, resulting in match score falling in this range. We also conducte calculations similar to those shown in the top graph of Figure \ref{fig:FAR_dis}, but this time we normalize the maximum amplitude to $5\times10^{-21}$. We repeat the same process for both simulated data with injected waveforms and simulated data without injected waveforms. As a result, we obtain the distribution of FAPR, which is illustrated in bottom of Figure \ref{fig:FAR_dis}. In the normalized case, when we set the threshold $\eta=0.75$, FAPR=$4\times10^{-3}$. We observe that, compared to the scenario where the GW source is located at 5 kpc and 10 kpc, the match scores of the data injected with GW signals are generally above the threshold. This result is attributed to our normalization of GW signals, effectively placing some waveform sources at closer distances. This adjustment facilitates better reconstruction of these GW waveforms. We also find that a small portion of the waveforms in Table \ref{tab:waveforms} could not be successfully reconstructed. The main reason is that these waveform having simpler temporal structures compared to the N1 waveform. As a result, during the calculation of the match score, the signal components of these waveforms carry lower weights, leading to the failure in reaching our predefined threshold. From the results in Figure \ref{fig:FAR_dis}, it can be concluded that when the GW source is closer and the GW signal is stronger, we have greater confidence in confirming that the reconstructed waveform is indeed a GW signal rather than noise.

\section{Discussion}
\label{sec:dis}

Extracting the GWs from CCSN needs a more complicated technologies as its time-frequency characteristics are not very intuitive as other common GW sources (such as compact binary coalescences). Therefore, in this work, we attempte to extract and reconstruct the GW waveform of CCSN by introducing EEMD. We evaluate the performance of EEMD based on a library of simulated CCSN gravitational waveforms (shown in Table \ref{tab:waveforms}). And we use match score to evaluate the differences between the reconstructed waveform and the injected waveform. Finally, we also perform an analysis to explore the probability that whether the extracted gravitational wave is a true signal.

We test a large amount of simulated data with GW sources at 5 kpc and 10 kpc, respectively, and use the match score to measure how well the reconstructed waveform matches the injected waveform. It is found that only the sum of the first six IMF decomposed by the EEMD method needs to be used as the reconstructed waveform. In addition, we also normalize the maximum amplitude of each waveform to $5\times10^{-21}$, and the components required to reconstruct the waveform are consistent with previous conclusions. The results are shown in Figure \ref{fig:Match}. Next, we analyze how far away the GW signal can be extracted and reconstructed by the EEMD method. We set the threshold of the match score as 0.75, and found that the distance at which the GW signal can be reconstructed is about 36 kpc, which is smaller than the distance that can be detected by aLIGO. This also means that in our study, the reconstruction distance is not equivalent to the detection distance. The results are shown in Figure \ref{fig:dis}. Finally, to assess the probability that our reconstructed waveforms represent real GW signals rather than noise, we first calculate the FAPR for the four waveforms at 5 kpc and 10 kpc, and the results showed that the FAPR was less than $2.5\times10^{-3}$, confirming the effectiveness and stability of the EEMD method. The results are shown in Figure \ref{fig:FAR_dis_case}. Next, we apply the same analysis to all waveforms in Table \ref{tab:waveforms}, calculating the FAPR at 5 kpc and 10 kpc, resulting in FAPR values of $6\times10^{-3}$ and $1\times10^{-2}$, respectively. To eliminate the influence of amplitude, we normalize the maximum amplitude of the injected GW signal to $5\times10^{-21}$, resulting in an FAPR of $4\times10^{-3}$. Figure \ref{fig:FAR_dis} present the corresponding results. Due to the detector response and relatively low amplitudes of some GW waveforms, approximately half of the waveforms cannot be successfully reconstructed at 5 kpc and 10 kpc. Additionally, from the bottom graph of Figure \ref{fig:FAR_dis}, we observed that a small portion of signals in Table \ref{tab:waveforms} could not be successfully reconstructed, even when the maximum of the amplitude of the waveform is normalized to $5\times10^{-21}$.

Comparing EEMD method with wavelets, the advantage of EEMD lies in its ability to decompose data as an adaptive process without the need for a basis set (e.g. \cite{Son2021JKPS}). For many signals whose mechanisms are not well understood, EEMD can avoid significant discrepancies between the reconstructed waveform and the true signal caused by an incomplete basis set. However, EEMD method may not effectively separate signals and noise when their frequencies are close. In recent years, the EEMD method has been applied in various fields of astronomy. For instance, it has been used for the analyzing of GW signals generated by the merger of compact binary stars or by Supernovae \citep{Hu2022ApJ}, as well as analyzing quasi-periodic oscillations \citep{Yu2023arXiv}. In future work, there is potential for further improvements to the EEMD method. For instance, when selecting the number of IMFs for reconstructing the waveform, we can opt for a more flexible choice based on some other criteria (e.g. the maximum amplitude of each IMF), rather than strictly summing the first six IMFs. We can explore using the EEMD method for the classification of GW signals generated by CCSN, which would help in confirming their explosion mechanism. Additionally, we can continue to search for better methods to extract GW signals. As future gravitational-wave detectors become more and more sensitive, such as Einstein Telescope \citep{Punturo_2010CQGra} and Cosmic Explorer \citep{Abbott2017CQGra}, it will be more and more possible to detect GW signals generated by CCSN.


\section*{Acknowledgements}
We thank the
anonymous referees for valuable comments and suggestions that
helped us to improve the manuscript.
We thank Jade Powell and Man Leong Chan for offering SN waveforms and helpful discussions.
Y. Yuan thanks Li-Lan Yang,  Lin Lan, Yu-Feng Li, Zhi-Qiang You, and Shao-qi Hou for their discussion. This work is supported by National Key R$\&$D Program of China (2020YFC2201400),  the National Natural Science Foundation of China (grant No. 11922303,11922301, 42230207), the Fundamental Research Funds for the Central Universities, Wuhan University (grant No. 2042022kf1182), the Joint Funds of the National Natural Science Foundation of China, grant number U2039205, and the Fundamental Research Funds for the Central Universities, China University of Geosciences (Wuhan) with No. G1323523064.  H. L\"{u} is supported by  the Guangxi Science Foundation the National (grant No. 2023GXNSFDA026007), and the Program of Bagui Scholars Program (LHJ). X. Fan is supported by  Hubei province Natural Science Fund for the Distinguished Young Scholars (2019CFA052).

\section*{Data AVAILABILITY}

This theoretical study did not generate any new data.



\bibliographystyle{mnras}
\bibliography{references} 


\bsp	
\label{lastpage}
\end{document}